# Development of theoretical frameworks in neuroscience: a pressing need in a sea of data

Horacio G. Rotstein[1] and Fidel Santamaria[2]

1 Federated Department of Biological Sciences, New Jersey Institute of Technology and Rutgers University, Newark, NJ 07102  horacio@njit.edu

2 Department of Neuroscience, Developmental and Regenerative Biology and Brain Health Consortium, The University of Texas at San Antonio, San Antonio, TX 78249  fidel.santamaria@utsa.edu

## Abstract

Neuroscience is undergoing dramatic progress because of the vast data streams derived from the new technologies product of the BRAIN initiative and other enterprises. As any other scientific field, neuroscience benefits from having clear definitions of its theoretical components and their interactions. This allows generating theories that integrate knowledge, provide mechanistic insights, and predict results under new experimental conditions. However, theoretical neuroscience is a heterogeneous field that has not yet agreed on how to build theories or whether it is desirable to have an overarching theory or whether theories are simply tools to understand the brain. Here we advocate for the need of developing theoretical frameworks as a basis of generating common theoretical structures. We enumerate the elements of theoretical frameworks we deem necessary for any theory in neuroscience. In particular, we address the notions of paradigms, models, and scales of organizations. We then identify areas with pressing needs to develop brain theories: integration of statistical and dynamic approaches; multi-scale integration; coding; and interpretability in the context of Artificial Intelligence. We also point out that future theoretical frameworks would benefit from the incorporation of the principles of Evolution as a fundamental structure rather than purely mathematical or engineering principles. Rather than providing definite answers, the objective of this paper is to serve as an initial and succinct presentation of these topics to encourage discussion and further in depth development of each topic.

# Introduction

There are three primary objectives of theoretical neuroscience: to provide quantitative testable predictions of brain function, to explain neuronal mechanisms in healthy and diseased brains and to abstract computational strategies to develop better engineering tools [1-4]. Achieving these goals requires developing theoretical approaches that encompass the shared assumptions of models developed for different species and scales [5-8]. If this were in fact possible, then experimental neuroscience would benefit from having access to theories that can take experimental data points obtained at one scale of organization and species and apply them to other scales and species. Thus, it is necessary to examine the present existing theoretical frameworks and determine whether they point to one or several theoretical approaches, what new frameworks need to be developed, and how to link them. It is no less important to clarify what is a theoretical framework, how frameworks are different from theories and models, and how different integrated modeling approaches produce a cohesive body of knowledge. These frameworks should build on the existing conceptual structure of the field and should not be compartmentalized, but must span across scales (e.g. spatial, temporal), levels of organization (e.g. subcellular, cellular, microcircuit, complex networks; systems; species), and theoretical sub-disciplines [9].

Traditionally, models and theories in neuroscience address a specific problem or data set [5,10]. In addition, theories have been founded on using metaphors (e.g., brain computation, the neural code)[11] and more metaphors to address these "primary" ones (e.g., brain-as-a-computer, information processing, and trajectories in the phase-space diagrams)[12,13]. However, the field would benefit from theories and models that integrate the vast data streams derived from the new technologies product of the BRAIN initiative to provide a mechanistic link between the activity of the nervous system and behavior [14]. Furthermore, theoretical frameworks need not only to integrate spatial and temporal scales, but also across biological scales of organization [15-18]. In order to achieve this, models and theories require a mathematical foundation for generalization and translation [19]. Generalization allows addressing larger problems or data sets. Translation refers to the ability of a theory developed at one scale, system, function, and species to be applied to another.

We have been involved in a number of activities whose goals are to bring together the theoretical neuroscience community and discuss the structure of the state-of-the-art, and identify fields within neuroscience that could benefit from developing frameworks that cross scales and established disciplines [20,21]. Some of these activities have resulted in publications on specific topics[21]. However, we believe there is a need to provide a succinct overview of the properties of neuroscience theories and areas to focus with a vision to the future. As such, the first part of this manuscript presents definitions of theoretical frameworks, paradigms, models, and scales of organization. We believe that there is a need to clarify each of these elements to develop and compare complementary or competing theories, such as in a more extensive recent paper. In the second part of the paper, we discuss five areas that require attention. In one of those topics, we emphasize that the theory of evolution should be included in the theoretical frameworks of neuroscience.

# Theoretical frameworks and their properties applied to neuroscience

We make use of the concept of theoretical frameworks to provide a parsimonious way to communicate and develop neuroscience theories. We identify the basic parts of theoretical frameworks applicable to any neuroscience field. In particular, we pay attention to the concept of scales of integration, which are taking more importance as large streams of data sets become available because of the technologies developed by the BRAIN initiative. Our objective in this section is to be succinct and to provide an initial introduction of the issues to promote conversation in the field and further development of these ideas.

## Frameworks

In a recent paper that resulted from our previous organizational efforts[21], a theoretical framework is a "language within which explanations are proposed." From this point of view, a framework is a set of theories believed to be fundamental that support a research program, creating a conceptual structure from which more theories are developed. Some other disciplines are keen in defining theoretical frameworks[22-26]. While there are variations, a summary of such proposals say that a theoretical framework consists of a theory or theories; their paradigms, both theoretical and experimental; and the fundamental concepts that identify them. A framework also includes the relationships and interactions among all these elements [27]. As mentioned before, it is not our intention on arguing on how to define theoretical frameworks for theoretical neuroscience. However, any definition should be useful to develop theories that integrate information across scales and across species. In this section, we will briefly discuss, what we argue, are the fundamental theoretical frameworks in neuroscience.

A first one is **dynamics**. In this case, computations are trajectories in phase-plane diagrams of neuronal activity (state) variables, such as membrane potential or firing rate. Coding is the transformation of a stimulus, static or dynamic, discrete or continuous, into a particular trajectory. Learning and memory are processes that shape these trajectories[28-30].

A **second** framework is statistics. In this case, neurons are stochastic agents that encode noisy signals. Memory, learning, and plasticity, take place in, mostly, a Bayesian inference setting, including causality, where there are priors of expected statistical models of the environment. The neuronal code is noisy and in need of networks that average or poll the responses of the system[31].

Engineering and machine learning concepts constitute a **third** framework. On one side of this group, there is an intrinsic assumption of a von Neumann type of arrangement of brain computation where there are modules that algorithmically parse and process information. Each module has a main task, which could interact with other modules depending on environmental conditions or learning[5,32]. The process is mostly hierarchical, providing higher levels of abstraction from the stimulus. Computational psychiatry is a good example of this approach[33,34]. In contrast, there is the non-von Neumann approach using neural networks[35,36]. However, hierarchical or deep neural networks theories have similarities with the compartmentalized approach[37,38].

While all these conceptual ideas on the structure of neuroscience theories are of importance, we could also adopt a different approach to move the field forward. Along these lines, a recent paper, the result of one of our previous organizational activities, took the pragmatic view to describe the function of theory and models in neuroscience [21]. Instead of proposing an overarching theory, it proposed to organize knowledge in terms of context-dependent frameworks, theories, and models, and to use three categories for each of them: descriptive, mechanistic, and normative. The descriptive category consists of phenomenological approaches to neuronal processes; mechanistic theories aim to describe the interactions of a number of components of the nervous systems; and normative theories use the biological function of a system under the constraints of an objective function.

Clearly, our discussion and classification of the different frameworks is a simplification and it is possible to find works that span these different perceptual approaches. However, it is important to identify and spell out their fundamental assumptions in order to be able to identify areas to develop. In addition, different framework approaches are not necessarily mutually exclusive, and can in fact be complementary.

# Paradigms

Theoretical frameworks allow organizing high-level problems by assuming relations among different components of neuroscience. It is at this level where we can use a set of paradigms, within a framework, to study the nervous system. A paradigm is a conceptual general view that consists of theories, classic experiments, and trusted methods [39]. The neuron as the basic block of the nervous system, synaptic plasticity as a cellular mechanism of learning and memory, the nervous system complexity supports its flexibility, are a few of these paradigms. These paradigms have been extremely useful in simplifying the study of the brain and allowing us to gain insight into neuronal function. However, how are the large multi-scale data expected from the technological advancements of the BRAIN initiative affect those paradigms?

# Models

While a paradigm is a statement of fundamental properties, a model is an abstraction (see discussion in [40]). Models are tools that contribute to our understanding of a problem and the interpretation of data, and their level of complexity satisfies these goals [41]. However, in many cases, the roles are reversed, and models distortedly became the foundation of a theory (e.g., the leaky integrate-and-fire model). Basing theory on oversimplified models distracts from advancing actual theory and forces researchers to design experiments constrained by the models. Model-based theory also results in producing theories that correspond to a specific spatial or temporal scale, a specific level of organization, or a specific experimental situation. The collection of such theories results in difficult-to-integrate higher order, more inclusive theories. In this respect, it is important to question if efforts should be devoted to this type of integrated frameworks or, alternatively, develop frameworks for the development of the necessary theories across scales.

## Scales of organization

The influence of physics and engineering in neuroscience has shaped the compartmentalized focus of theories [42-45]. As such, theories aim to explain brain phenomena at separable physical or temporal scales. For example, the impulse-response strategy to characterize everything, from the opening of an ion channel to mapping the motor cortex in primates, assumes that it is enough to characterize the response of the system to a short stimulus to understand the embedding macro-systems [46,47]. Building a theory on only this type of oversimplified assumptions proves to be of limited predictive value when, for example, the stimulus is longer, is presented in different contexts, or the response is affected by nonlinearities [48]. At this point, it is relevant to bring into the discussion the concept of **biological scales of organization**[49]. These levels of organization are not just separable physical or temporal scales, but describe interactions among elements and make relationships to their isolated forms (networks vs single neurons)[50,51]. Integrated frameworks should include mechanisms of looking at specific levels of organization by considering the effect of their sub- and supra-levels in ways that are consistent across scales and their biological function. Here, the use of Artificial Intelligence (AI) for automatic detection of complex behavior, together with electrophysiology and modeling, and additional tools to resolve degeneracies could provide the experimental foundation to determine biological scales by function and behavior [52].

# Areas of neuroscience that require theoretical development

Here we present an initial discussion of different topics that, we believe, will be of importance in the near future. As in the previous section, it is not our intention to have an exhaustive discussion of each one of them, but instead to present them in a succinct form to inspire debate.

## Integration of dynamics, statistics and other mathematical approaches

The dynamical and statistical neuroscience communities have approached problem solving and the generation of theories and models from different perspectives, using different tools and different languages [53-56]. While there are overlaps between the different approaches a conceptual separation between the ways the two fields address theoretical neuroscience questions remains prominent. For example, the leaky integrate-and-fire model and the generalized linear model are largely considered to be equivalent. How do the equivalences between primarily dynamics and statistical models extend to more complex systems? Statistical tools are often calibrated or validated by using the so-called ground truth models. Can this approach be extended to provide interpretability to the results of using these tools in combination with experimental/observational data? How can a mathematical integrated approach further incorporate other fields of mathematics that are becoming increasingly useful in theoretical neuroscience such as topology and geometry? Finally, how these combined approaches can contribute to develop tools to establish causal effects among neuronal processes, understand biophysical and dynamic mechanisms, and resolve degeneracies present in neuronal systems?

## Multi-scale challenges

From the beginning of modern neuroscience, it has been necessary to break down the tasks of the nervous system into elements tractable by the experimental and theoretical tools available at the time. Thus, there was no pressing need for a multi-scale approach. In this context, the mean-field assumption is a widely used approach to study the effects of other scales on a specific problem. The fundamental assumption of this process is that the activity that is being averaged out can be represented as a homogenous population [57-61]. The assumptions of what constitutes a homogeneous population have to be carefully understood before simplifying their description. Novel approaches are needed to consider the effects discussed above of the sub- and supra-levels of organization with respect to the one considered in ways that are consistent across scales and preserve the necessary details to ensure the dynamics across scales are preserved.

## Coding

There is no single definition of coding [62-66]. From a correlative point of view, coding, by either spike timing or firing rate strategies, is a static process characterized by mutual information. A second form of coding is by the biophysical decoding properties of the downstream receivers. Changes in neuronal excitability due to neuromodulatory or intrinsic processes requires the recalculation of the mutual information. This neuron-centric approach creates conundrums in which it is not possible to know for a receiver when a pre-synaptic neuron has changed coding strategies. Thus, it is necessary to define coding and establish how different notions of coding interact at multiple scales, from synapse to system.

## Mechanistic interpretability of artificial intelligence and machine learning in neuroscience

Experimental and computational neuroscientists are increasingly using artificial intelligence (AI), particularly machine learning (ML) tools, to analyze large data sets with for a number of purposes, including data processing, understanding relationships between variables, uncovering hidden structure of neuronal activity, data classification, data clustering, and hypothesis testing[67,68] [69-71]. Complemented with additional tools, these primarily data-driven approaches can be functional for hypothesis generation, hypothesis testing and certain aspects of mechanistic interpretation of the data. However, the process of automatic discovery of neuronal mechanisms using AI, a concept known as explainable AI or XAI [71,72], as well as the dynamic properties of the systems that underlie neuronal processes is still underdeveloped. Overcoming these issues is crucial for the field to move forward and to produce theories that are not only interpretable, but also generalizable beyond the limits of the data sets used to produce them. Moving forward requires the development of novel conceptual ideas and algorithms, and platforms that allow these ideas to be implemented. For example, by taking inspiration from the shortcomings in other fields[73], we could use XAI for the testing and developing of frameworks, theories and models. Since we are at the beginning of incorporating these technologies to research workflows, it is important to train a new generation of investigators that experience this as the standard approach.

## Evolution as a framework of theoretical neuroscience

The brain is the result of evolution. However, there are multiple ways of using Evolution to study the brain. In possibly the most common form, Evolution studies how the shapes and structures of the nervous system relate among organisms. A second branch aims to understand brain function in relation to behavior, mainly mating and survival, at evolutionary scales. But, Evolution is also an engineering metaphor to develop training and learning strategies that could explain the diversity of the brain's computational strategies.

In contrast to the more common uses, we propose using Evolution to study the principles of brain computation. We could use the evolutionary basic concepts of allopatric, sympatric, adaptive radiation, hybrid zones, and the Hardy-Weinberg principle of equilibrium to study not only brain evolution but also neuronal computation[74,75]. We could do this at two scales. The first one is the study of behavior across species using evolution to develop computational theories of the brain. A fascinating example is on the evolution of vocal communication and speech. Recent work provides strong evidence that vocal communication can be described with all the traits of evolution, including convergent evolution [76]. If indeed, evolution explains vocal communication then we propose that each implementation of this behavior represent a ***computational phenotype***. Each computational phenotype vary by the presence of a ***computational allele***, which could be a particular network implementation or coding strategy. In this context, we expect preservation and variation of computational strategies and their neuronal substrates across related species. Another area of study that should be taken into account is learning and memory of organisms in hybrid zones [77]. Hybrids could have different learning strategies, possibly putting them in a disadvantage to the parental diverging populations; thus, contributing to the generation of species.

In the second scale, we look at computational phenotypes within a single species. Here, the diversity of computational alleles could result in adaptive radiation, where organisms express a given allele in higher frequencies. This adaptive radiation could take place in populations that are in the same (sympatric) or different (allopatric) environments. Computational alleles provide, just as in any other instances of alleles, flexibility to the survival of the population in case of environmental changes. In contrast to ideas that species in a stable environment develop a unique, engineering inspired, computational solution, we expect to have multiple computational strategies. Each computational allele could be optimal under different environmental conditions. Under an engineering framework, this would mean that some organisms are less optimal. However, under an evolutionary framework, this means that the population is retaining computational strategies for the survival of the species. Furthermore, we expect that the distribution of computational strategies have a stable distribution under similar environmental conditions, and for this distribution to change under environmental manipulation, as described by the Hardy-Weinberg principle of equilibrium. Thus, evolutionary principles suggest that we should find a distribution of computational strategies to solve particular problems or processes in genetically similar populations under similar environmental conditions.

Clearly, an open question is how to fit all the other frameworks under an evolutionary umbrella. In principle, under evolution, organisms could benefit from the presence of dynamical, statistical or machine learning strategies, all present and varying at population levels in order to maximize survival of the species.

## Training the new generation and next steps

The primary current model to train engineers, mathematicians, or physicists that enter into a neuroscience doctoral program is through rotations in, hopefully, experimental laboratories, taking neuroscience classes, or participating in dedicated summer schools. For postdoctoral researchers the training is far less homogeneous, mostly consisting on participating in seminars and dynamics within the mentor's laboratory. Intrinsically, this training strategy assumes that it is only necessary to add some neuroscience topical training for students with these backgrounds to understand how the brain works. To some extent, the training of these new students is a closed-loop circuit, given that several of the most used textbooks to teach theoretical neuroscience are by physicists or mathematicians.

We tried to address this issue as part of a workshop we organized, supported by NSF [78]. We mixed experimentalist and computational researchers, we purposely integrated the active participation of graduate students and postdocs [79,80], and we started the conversation well in advance of the meeting itself. Our aim was to integrate students and postdocs in the discussion rather than being passive listeners. There are other activities, such as the NIH supported Theories Modeling and Methods workgroup. Recent efforts within this group aimed to create synergies between theory and experimentally led groups [81]. These activities centered in exchanging data and models. We envision a future where students and postdocs benefit from adding this type of training activities.

In fact, we advocate that training the next generation of theoretical neuroscientists should incorporate additional components that put the traditional training in the context of a more general form of "biological thinking". Training should include an understanding of theoretical biology in the forms of evolution and ecology in the context of the development and function of the nervous system. Complementary, it would be beneficial for theoretical neuroscientists to understand systems biology and, more generally, the subcellular processes of the brain. This objective could make use of already present training programs. However, it would also be useful to develop multidisciplinary textbooks that integrate the engineering/mathematical traditional side of theoretical neuroscience with theoretical biology, written in collaboration with experts in the field.

Support:

F.S. NSF-IOS 1516648 and NIH NIMH-NIBIB R01EB026939

H.G.R. NSF-DBI 1820631, NSF-CRCNS-1608077, NSF-IOS-2002863